\documentstyle[12pt]{article}

\def\bq{\begin{quotation}}
\def\eq{\end{quotation}}

\def\fnote#1#2 {\begingroup \def \thefootnote {#1}
\footnote{#2}\addtocounter{footnote}{-1}\endgroup}

\newcommand{\support} {This work is supported in part by U.S. Department of
Energy Contract No.  DE-AC02-76ER13065.}

\newcommand{\myname} {\vspace{0.5in}
                                  \begin{center}Zhu Yang\\
                                  \vspace{0.2in}
                                  Department of Physics and Astronomy\\
                                  University of Rochester\\
                                  Rochester, NY 14627\\
                                  \vspace{0.5in}
                                 Abstract\\
                                  \vspace{0.2in}
                                \end{center}}

\newcommand{\pagenumber}{\pagestyle{plain}\setcounter{page}{1}}

\def\d{\delta}

\def\m{\mu}

\def\s{\sigma}

\def\raisenot{\raise .5mm\hbox{/}}
\newcommand{\notpa}{\hbox{{$\partial$}\kern-.54em\hbox{\raisenot}}}
\def\notp{\ \hbox{{$p$}\kern-.43em\hbox{/}}}
\def\notq{\ \hbox{{$q$}\kern-.47em\hbox{/}}}
\def\notk{\ \hbox{{$k$}\kern-.47em\hbox{/}}}
\def\notA{\ \hbox{{$A$}\kern-.47em\hbox{/}}}
\def\nota{\ \hbox{{$a$}\kern-.47em\hbox{/}}}
\def\notb{\ \hbox{{$b$}\kern-.47em\hbox{/}}}

\begin {document}
\baselineskip=24pt

\pagestyle{empty}

\begin{flushright}
hepth@xxx/9202078
UR-1251\\
ER-13065-701\\
\end{flushright}
\vspace{0.5in}
\begin{center}
{\Large A Possible Black Hole Background in c=1 Matrix Model}
\end{center}
\myname
We propose a new space-time interpretation for c=1 matrix model with
potential $V(x)=-x^{2}/2-\m^{2}/2x^{2}$. It is argued that this
particular potential corresponds to a black hole background. Some
related issues are discussed.
\newpage
\pagenumber
c=1 matrix model turns out to have very rich  structure. It possesses, for
example, a propagating degree of freedom, a non-trivial $S$-matrix, discrete
states, a large symmetry algebra $w_{\infty}$, and so on. For
a review, see \cite{igor}.
All these have their corresponding counterparts
in the continuum  Liouville theory as well. Now viewed as a critical $D=2$
string theory, the continuum theory also has  a black
hole solution \cite{elit},
which has not been seen in matrix models. A priori, there are problems.
If a black hole radiates, it should be in the physical spectrum. On the
other hand, the matrix model is already unitary. Closely related is the fact
that the matrix model does not introduce the dilaton and metric degrees of
freedom explicitly, so it is not clear how to study the back reaction,
if there is one.
It is however recently argued that the Euclidean black hole mass is
a superselection parameter and does not fluctuate \cite{nath}.
This gives us some
hope that the black hole may be represented by a specific one-body Hamiltonian
of the matrix model.


There is another reason to look for different Hamiltonian than the usual
inverted harmonic oscillator. Compare the work in matrix model approach and
the Liouville approach, we can make identification of special
operators in the two theories \cite{dg}.
It turns out the ``correct" Liouville
dressing of a primary matter operator corresponds to a polynomial with
time dependence in the matrix model. An infinitesimal black hole,
being ``wrongly" dressed, would correspond to a potential
of negative power in the matrix model. Such a potential is a
relevant perturbation in the sense that it alters drastically the
critical behavior. So the correct way to solve the problem is to go beyond
the usual perturbation theory and it is interesting to study this as a part
of larger program: investigate the relevant perturbations and
find out their physical interpretations \footnote{Multicritical points with
the potential $x^{n}$ ($n>0$) have been considered by Gross and Miljkovic
\cite{bipz}. See  also \cite{das} for
different approaches.}.

We first briefly
review what is known about the black hole solution in the continuum
theory. Then we consider general coupling of special states in the matrix
model, and finally specialize to the possible black hole background.

The Euclidean black hole can be summarized by the following metric
and dilaton \cite{elit}
\begin{eqnarray}
ds^{2} &=& (1-Me^{-\phi})dt^{2} + (1-Me^{-\phi})^{-1}d\phi^{2} \nonumber \\
\Phi &=& \phi.
\end{eqnarray}
Here we have chosen a coordinate system where dilaton is identified with the
spatial coordinate, and $M$ is the black hole mass.
The world sheet action can be written as (in a flat world sheet background)
\begin{equation}
S={\frac{1}{2}}\int d^{2}\s [{\frac{1}{ 1-Me^{-\phi}}}\partial_{z}t\partial_{
\bar{z}}t+(1-Me^{-\phi})\partial_{z}\phi\partial_{\bar{z}}\phi].
\end{equation}
{}From (2), if we use the minisuperspace quantization, which is known to
be exact for $c\le 1$ in the Liouville background \cite{greg}, we have
the following
Wheeler-de-Witt (WdW) wave equation \cite{divi}
\begin{equation}
{\frac{1}{1-Me^{-\phi}}}{\frac{\partial^{2}}{\partial t^{2}}}\Psi
+{\frac{\partial}{\partial \phi}}(1-M e^{-\phi}) {\frac{\partial}
{\partial \phi}} \Psi =0.
\end{equation}
We will find (1) and (3) follow from the matrix model as well.

Now we turn to c=1 matrix model.
We will use the collective field theory as our starting point \cite{anta},
keeping in mind the underlying fermion picture \cite{bipz}.
This Thomas-Fermi approach has been advocated by Polchinski \cite{joe}.
The collective field seems to describe the tachyon dynamics. Its precise
relation with the tachyon in the Liouville theory is however not
clear. There may be some non-local field redefinition between the
two \cite{dave}.
Actually the question is more general: it is not clear how to relate
the spacetime picture in the two approaches. For example, the WdW equation
derived from the minisuperspace corresponds to the Laplace transformed
equation of motion of the linear fluctuation of the collective
field \cite{greg},
which certainly points to a subtle relationship between the two \cite{gmns}.

In what follows we find that the correspondence of our matrix
model with the black hole is much more straightforward, a point
related perhaps with the dual relation between the black hole and the
Liouville theory discovered in \cite{emil}.

1+1 dimensional string from $c=1$ matrix model is described by $N$ fermions
with Hamiltonian,
\begin{equation}
H_{F}=\int^{\infty}_{-\infty} dx [{\frac{1}{2}} \partial_{x}\psi^{\dagger}
\partial_{x}\psi + V(x) \psi^{\dagger}\psi],
\end{equation}
whose bosonized form is the collective field theory with the Hamiltonian
\begin{equation}
H_{B}={\frac{1}{2\pi}}\int^{\infty}_{-\infty}[{(\frac{p_{+}^{3}}{6}}+
p_{+}V)-({\frac{p_{-}^{3}}{6}}+p_{-}V)].
\end{equation}
In the boson form,
\begin{eqnarray}
p_{\pm} &=& \pi_{\xi}\pm \partial_{x}\xi \\
\lbrace\pi_{\xi}(x,t), \xi(x^{\prime},t)\rbrace_{P.B.} &=&
2\pi \partial_{x}\delta (x-x^{\prime}).
\end{eqnarray}
Since $\partial_{x}\xi$ is the density of states, it must be positive
semi-definite. So we are dealing with a fluid field theory.
In the usual double scaling limit, $V(x)$ is an inverted harmonic oscillator,
\begin{equation}
V_{1}(x)=-{\frac{1}{2}}x^{2}.
\end{equation}
We will discuss another potential,
\begin{equation}
V_{2}(x)=-{\frac{1}{2}}x^{2}-{\frac{\mu^{2}}{2x^{2}}},
\end{equation}
which coincides with $V_{1}(x)$ when $|x|\rightarrow\infty$. It can be
considered as a different fine-tuning of the critical potential.

The equations of motion of $p_{\pm}$ are
\begin{equation}
\partial_{t}p_{\pm}=-V^{\prime}(x)-p_{\pm}\partial_{x}p_{\pm}.
\end{equation}
A natural starting point to solve the theory would be to expand
$p_{\pm}$ around static background  $\bar{p}_{\pm}$ given by
\begin{equation}
V^{\prime}(x)+\bar{p}\partial_{x}\bar{p}=0,
\end{equation}
where
\begin{equation}
\bar{p}_{+}=-\bar{p}_{-}\equiv \bar{p} \ge 0.
\end{equation}
We are interested in the case when the fermi surface is right at the top of
the potential. It corresponds to zero cosmological constant on the world
sheet in continuum language, an ansatz used in \cite{elit}.


To figure out the space-time metric of the background (11), we must look for
excitations around it, following \cite{anta}.
Let
\begin{eqnarray}
\xi &=& \bar{\xi}+\delta \xi,\\ \nonumber
p_{\pm} &=& \bar{p}_{\pm}+\delta p_{\pm},
\end{eqnarray}
we have for the action $S$,
\begin{eqnarray}
S &=& \bar{S} + \int dx\,dt\, \lbrace\pi_{\xi}\delta\dot{\xi}
 -[{\frac{1}{2}}\bar{p}(\delta p_{+}^{2}-\delta p_{-}^{2})
+{\frac{1}{6}}(\delta p_{+}^{3}-\delta p_{-}^{3})]\rbrace \\ \nonumber
 &=& \bar{S} + \int dx\,dt\, \lbrace\pi_{\xi}\delta\dot{\xi}
 -[{\frac{1}{2}}\bar{p}(\pi_{\xi}^{2}+ (\partial_{x}\delta\xi)^{2})
+{\frac{1}{6}}(\delta p_{+}^{3}-\delta p_{-}^{3})]\rbrace,
\end{eqnarray}
where we have used the definition of $p_{\pm}$ in the second line.
Looking only at the quadratic piece of (14), we can see that it corresponds
to a massless scalar field propagating in an external metric background.
In order to make connection with the continuum theory, it is more
convenient to use $\phi = \ln (-x)$ as the spatial coordinate and
make a scaling $\pi_{\xi}\rightarrow \pi_{\xi}/x$. With this done,
(14) becomes
\begin{equation}
S= \bar{S} + \int dx\,dt\, \lbrace\pi_{\xi}\delta\dot{\xi}
 -[{\frac{1}{2}}{\frac{\bar{p}}{x}}
(\pi_{\xi}^{2}+ (\partial_{\phi}\delta\xi)^{2})
+{\frac{1}{6x^{2}}}(\delta p_{+}^{3}-\delta p_{-}^{3})]\rbrace.
\end{equation}
The string coupling constant is obviously $\exp(-2\phi)$. So $\phi$ can
be thought as the dilaton.

Before we plunge into the details, let us remark on the
discrete state in matrix model. Consider
adding  the following special operator in
the Hamiltonian,
\begin{equation}
O_{mnl}=\int dx (p^{m}_{+}-p^{m}_{-})x^{n} e^{ilt},
\end{equation}
where $m,n$ and $l$ are integers.
By the way, although one can construct $w_{\infty}$ generators from the above
construction for both $p_{+}$ and $p_{-}$, the underlying fermion picture
allows only the diagonal $w_{\infty}$ as the dynamical algebra, a fact also
well-understood in the Liouville theory.
Back to (15), note that not
every operator is independent, however\footnote{The
following remark belongs to J. Polchinski.}. In the classical theory,
an perturbation of (8) that can been written
as a total time divergence is considered merely as
a canonical transformation, so it does not change the physics.
Since
\begin{equation}
{\frac{d}{dt}}O_{mnl}=il O_{mnl}+{\frac{mn}{m+1}}O_{m+1,n-1,l}+
m O_{m-1,n+1,l}.
\end{equation}
We can successively use (17) to relate different operators
and the independent operators are labeled by only two integers.
We can choose their form as, for example,
\begin{equation}
O_{ml}=\int dx (p_{+}-p_{-})x^{m} e^{ilt}.
\end{equation}
(18) is an irrelevant perturbation if $m>0$.
Now an infinitesimal black hole has the vertex operator
\begin{equation}
B=M(\partial_{z}t\partial_{\bar{z}}t-\partial_{z}\phi\partial_{\bar{z}}\phi)
e^{-2\phi}.
\end{equation}
In matrix model language, we propose the following operator be identified with
(19),
\begin{equation}
O_{2,0}=\int dx (p_{+}-p_{-})x^{-2}.
\end{equation}
This is a relevant perturbation, since it changes the critical behavior
quite a bit. A natural thing to resolve this seems to impose suitable boundary
conditions near $x=0$.  We will define the critical point to be
when the chemical potential reaches the maximal of the potential.
The boundary condition is such that fermion wave function is zero
at the maximal. This mimics the hard wall of the Euclidean
black hole.

Consider now the potential (9). The time independent background is given by
\begin{equation}
p= x-\mu/x.
\end{equation}
In (21) we have tuned chemical potential to be exactly at the top of the
potential. This makes sense because the black hole (1) is characterized by
one parameter $M$.
We will see shortly that $\mu$ labels the black hole mass.
The action (15) becomes
\begin{equation}
S=\bar{S}+ \int d\phi dt [\pi_{\xi}\dot{\xi}-{\frac{1}{2}}
(1-\m e^{-2\phi})( \pi_{\xi}^{2}+(\partial_{\phi}\xi)^{2})
+{\frac{1}{6}}e^{-2\phi}(\d p_{+}^{3}-\d p_{-}^{3})].
\end{equation}
We see that the linearized equation derived from (22) is exactly
(3) without any field
redefinition. Besides, the string coupling is obviously of the standard form
$\exp (-\phi)$.
So identifying $\phi$ with the dilaton background is very natural.
If we accept these, we conclude that (22) describes
$D=2$ string moving in a black hole background.

It is now straightforward to use the effective action to calculate the
scattering amplitudes. We leave it to a future work. Here we would like to
make some general remarks.

1. Why is that for the cosmological constant background, $\phi$ is not
identified with the Liouville field zero mode, whereas for the
black hole it is?

2. Consequently, defining metric, dilaton and tachyon background is tricky.
This fact has been appreciated by many authors. It seems that our
work adds one more subtlety: We modify the
Hamiltonian and make an explicit separation of the tachyon and
the metric backgrounds. It is not clear how to study a combined
tachyon and black hole background, since we make a direct identification
of the dilaton with $\ln (-x)$, yet the relation is more indirect for the
cosmological constant background. It is very important to understand this
issue.

3. A closely related problem is to find a $\s$-model action that reproduces
the exact  results for both cosmological background and black hole.
In \cite{ramy} it is proposed to ``covariantize" the inverse harmonic
oscillator collective field theory action. Due to the problem mentioned
above, it is not clear to what extent it can be valid for the both
backgrounds. This may be related to the field redefinition issue in
string theory.

4. The matrix model action seems to make sense in Minkowski space too. How
would this affect our understanding of the Minkowski black hole? Is it possible
that it is stable? Also, even in Euclidean space, there is no apparent
 reason to compactify $t$ in the matrix model. Clearly some deeper
understanding is needed, especially of the path integral measure problem.

5. Previously, an attempt has been made to study black holes in the usual
$c=1$ model matrix \cite{joe2}.
The idea is to study the dynamics of the formation and
subsequent disappearance of black hole due to tachyon fluctuation. It was
find that the tachyon self-interaction is too strong.
Maybe this is the indication that the usual matrix model Hilbert space does not
include black hole. One must change the dynamics in order to see it.

6. What does a more  general potential $V(x)$ correspond?
Several recent studies \cite{witten,kleb}
find that the negative power operators act as derivatives
on a distribution--very singular objects. Some appropriate boundary
conditions are needed to make sense of them, as we have done
in this paper.

7. In order to completely confirm our proposal, we must
study the matrix model black hole is to examine correlation
functions and compare with the continuum theory. This is under study.

In conclusion, we have suggested a new space-time interpretation of
the c=1 matrix model. With a modified critical potential, a possible
black hole background emerges. Lots of questions remain to be answered.

\vspace{0.3in}
\begin{flushleft}
{\Large Acknowledgement}
\end{flushleft}
I would like to thank R. Brustein, A. Jevicki, D. Minic, and
especially J. Polchinski for various discussions on the related
subjects. \support

\end{document}